\documentclass[12pt]{iopart}
\usepackage{enumerate}
\usepackage{multicol}
\usepackage{graphics,graphicx}

\begin{document}

\title{Experimental and numerical characterisation of the turbulence in the Scrape-Off Layer of MAST}

\author{F. Militello$^1$, P. Tamain$^2$, W. Fundamenski$^3$, A. Kirk$^1$, V. Naulin$^4$, A.H. Nielsen$^4$ and the MAST team}
\address{$^1$ EURATOM/CCFE Fusion Association Culham Science Centre, Abingdon, Oxon, OX14 3DB, UK}
\address{$^2$ CEA-IRFM, F-13108 Saint-Paul-lez-Durance, France}
\address{$^3$ Imperial College of Science, Technology and Medicine, London, UK}
\address{$^4$ Association EURATOM- DTU, Technical University of Denmark, Department of Physics, DTU Ris$\o$ Campus, P.O. Box 49, 4000 Roskilde, Denmark}

\begin{abstract}

Numerical simulations of interchange turbulence in the Scrape-Off Layer are performed in a regime relevant for a specific L-mode MAST (Mega Ampere Spherical Tokamak) discharge. Such a discharge was diagnosed with a reciprocating arm equipped with a Gundestrup probe. A detailed comparison of the average and statistical properties of the simulated and experimental ion saturation current is performed. Good agreement is found in the time averaged radial profile, in the probability distribution functions (PDFs) and in qualitative features of the signals such as the shape, duration and separation of burst events. These results confirm the validity of the simple interchange model used and help to identify where it can be improved. Finally, the simulated data are used to assess the importance of the temperature fluctuations on plasma potential and radial velocity measurements acquired with Langmuir probes. It is shown that the correlation between the actual plasma quantities and the signal of the synthetic diagnostics is poor, suggesting that accurate measurements of the temperature fluctuations are needed in order to obtain reliable estimates of the perpendicular fluxes.   
 
\end{abstract}

\maketitle

\section{Introduction}

The exhaust of the plasma is a central problem for the design of the next generation magnetic fusion experiments and in the perspective of a reliable reactor  \cite{Loarte2007}. The narrow Scrape-Off Layer (SOL) at the plasma boundary regulates the outflow of energy and particles toward the solid materials at the divertor targets and at the first wall, which, without appropriate control, would not be able to withstand the massive energy required to trigger and maintain nuclear fusion. It is therefore crucial to reach a sufficiently detailed understanding of the physics of the SOL and, in particular, of the mechanisms that transport the energy within it. This challenge is further complicated by the exotic environment which provides the background of this problem. Relative to the core, the plasma boundary is cold, rarefied, highly non-adiabatic and non-linear, with a magnetic topology characterized by field lines that intercept solid targets \cite{Militello2011}. All these factors contribute to complicate the description of the SOL.

On the other hand, several experimental measurements in different tokamaks provide a comprehensive picture of this region of the plasma. Langmuir probes, fixed or installed on reciprocating arms, were the workhorse of these kind of measurements. The SOL was thus generally described as turbulent, with large fluctuations with respect to the background (especially in the density), highly intermittent and not conform to Gaussian statistics \cite{Zweben1985,Boedo2001,Garcia2007a,Garcia2007b}. 

The main features of these experimental observations were captured by an elegant theoretical interpretation which describes the perpendicular transport in the SOL as governed by the advection of plasma filaments (also called blobs) driven by interchange motion \cite{Sarazin1998,Krashenninikov2001,Garcia2004,Garcia2006b}. In addition to providing a qualitative explanation of the behaviour of the plasma boundary, this theoretical framework also proved to be reliable in quantitative comparisons with experimental data. In particular, salient features of the SOL in TCV (Tokamak a Configuration Variable) \cite{Garcia2007a,Garcia2007b,Garcia2006,Garcia2007} and JET (Joint European Torus) \cite{Fundamenski2007} were captured by the ESEL (Edge-Sol ELectrostatic) code \cite{Garcia2004,Garcia2005} (positive preliminary results were also obtained for MAST \cite{Tallents2009}). More recently, this code was used to characterize the operating regime of the spherical tokamak MAST \cite{Militello2012}. This analysis also offered an explanation for some observed dependencies between global plasma quantities (e.g. line averaged density) and SOL features (e.g. the density and temperature decay lengths). 

In this paper, we report on a comparison between a specific L-mode MAST pulse (\#21712) and a dedicated ESEL simulation. The first aim of this study is to retrospectively provide an experimental back up to the analysis in \cite{Militello2012}, which was performed for "typical" (but generic) MAST configurations. In this sense, the present manuscript can be seen as a companion paper to \cite{Militello2012}. Our investigation, however, is not limited to a pure code validation. Indeed, the detailed comparison against experimental data allows to identify when the theoretical model implemented in the code fails to deliver accurate matches. This allows us to clarify the boundaries and limits of applicability of the code to experimental interpretation and therefore prediction. In addition, our work can be seen as tool at the service of future theoretical developments, as improved models will require to heal the specific discrepancies observed.

Finally, as an application, we use the ESEL simulations to determine the quality of Langmuir probe measurements for MAST conditions. In particular, it was recently observed \cite{Gennrich2012,Nold2012}, that the correct inclusion of the electron temperature fluctuations is essential in order to relate the floating electrostatic potential (measured by the probe) to the actual plasma electrostatic potential (which is what we want to determine). An appropriate estimate of the latter quantity is essential as it is used to calculate the $\textbf{E}\times\textbf{B}$ velocity in the SOL and, consequently, the radial particle and energy flux. However, the measurement of the electron temperature in the SOL is a delicate procedure and it is often not available (while it is an easily readable output of every ESEL simulation). 

The paper is structured in the following way. The experimental set-up and the measurements of the ion saturation current used in our analysis are described in Section \ref{SecII}. Section \ref{SecIII} describes the physical model used and its implementation in the ESEL code. The experimental and numerical averaged profile and fluctuating components of the ion saturation current are introduced in Section \ref{SecIV} where the comparison between MAST data and ESEL simulations is discussed in detail. The effect of the temperature fluctuations on the Langmuir probe measurements is studied in Section \ref{SecV}. Finally a summary of the paper and our conclusions can be found in Section \ref{SecVI}.      
 
\section{Experimental measurements}\label{SecII}

MAST is a spherical tokamak with tight aspect ratio: its major and minor radii are $R=0.85m$ and $a=0.59m$ respectively, giving $\epsilon=a/R=0.64$. The pulse modelled in our analysis is the connected double null discharge \#21712, which was in an ohmic L-mode regime. It was characterized by a plasma current of $I_p=400 kA$ and a confining magnetic field of $B_T=0.4 T$ on the magnetic axis (the resulting safety factor is $q_{95}\approx 6.2$). During the flat top phase, the core electron temperature was $T_e\approx 650 eV$ and the line averaged density $\overline{n}\approx 1.7\cdot 10^{19} m^{-3}$. 

\begin{figure}
\includegraphics[height=9.5cm]{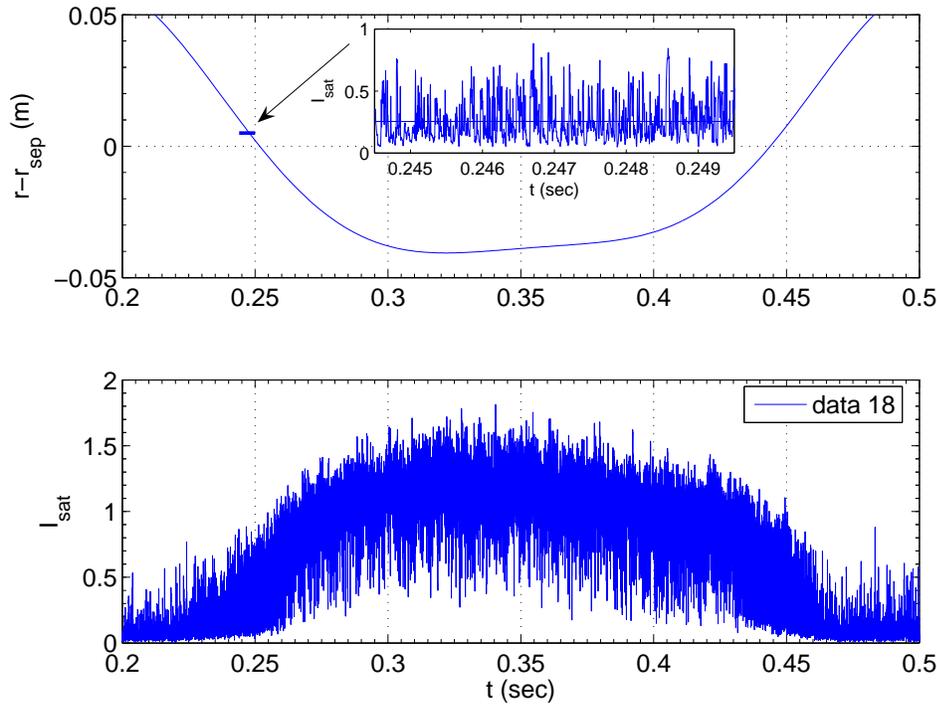} \caption{Upper frame: position of the top of the reciprocating probe as a function of time. Lower frame: total raw signal of the ion saturation current (A.U.) as a function of time. In the small frame: zoom of the raw $I_{sat}$ signal in one of the intervals (indicated by the arrow).}
\label{fig0}
\end{figure}

The measurements that we analysed were obtained with a Gundestrup probe \cite{MacLatchy1992} installed on a mid-plane reciprocating arm \cite{Yang2003}. The probe reciprocated inside the plasma and out during the flat-top phase, acquiring data at 500 kHz. Its radial position with respect to the separatrix as a function of time is shown in the upper frame of Fig.\ref{fig0}. Eight Pins are uniformly distributed around the probe's cylindrical head and were biased to $-200 V$ in order to measure the ion saturation current, $I_{sat}$ (a schematic of the probe can be found in Refs.\cite{Tamain2010,Higgins2012}). In our study we used the signal from Pin 1, which is the least affected by the disturbances induced in the plasma by the probe, see the lower frame of Fig.\ref{fig0}. Another three Pins come out of the probe's front plane and were used to measure the floating electrostatic potential, $\phi_{fl}$. 

During the reciprocation, the value of the plasma parameters remained approximatively constant and equal to the nominal values given above.
%between $0.3$ sec and $0.45$ sec, with small variations in the intervals $0.2-0.3$ sec and $0.45-0.5$ sec. 
In order to extract a radial dependence of the SOL features, we identified a set of radial positions around which most of our analysis was performed. We chose 21 of those, going from $r-r_{sep}=-0.04$ m to $r-r_{sep}=0.04$ m with a separation of $0.004$ m (note that positive values represent points in the SOL). The raw $I_{sat}$ signal was then divided into sub-signals lasting 5 ms and centred around the times at which the probe was intercepting the 21 points. This time window was chosen short enough so that the radial displacement of the probe around the nominal radial position was small ($\ll 1$ cm). In the case the probe was passing more than once through the same radial position a new interval was added in order to improve the statistics of the signal. We therefore ended up with 21 sub-signals, each one "characteristic" of a certain radial position.  

The systematic procedure described above allowed us to calculate an average value of $I_{sat}$ at each radial location and it also provided sub-signals of a sufficient length for a statistical analysis. Although some of the intervals thus identified overlap, this does not significantly affect our results, as we verified by performing a similar analysis with non-overlapping sub-signals (which is obviously less radially resolved).

\section{Physical and Numerical Model}\label{SecIII}

The physical model implemented in the ESEL code is a set of electrostatic drift-fluid equations solved in a 2D numerical domain that represents a portion of the drift plane localized in the outboard mid-plane of the tokamak. The model is entirely self-consistent in the plane perpendicular to the magnetic field, while the parallel dynamics is removed from the problem and replaced by \textit{ad hoc} loss terms. These, together with the collisional dissipation, are modelled using reasonable physical assumptions and are uniquely determined by the conditions at the edge of the plasma. In other words, the equations do not contain adjustable parameters that can be tuned in order to obtain a good matching with the experimental results. 

From a geometrical point of view, the code simulates three distinct plasma regions. Moving from the core to the wall, the first one is the \textit{edge}, a few centimeters inside the last closed flux surface (LCFS). Here no parallel losses are present due to the periodicity in the poloidal and toroidal directions. The next region is the \textit{SOL}, a few centimeters outside the LCFS (the magnetic connection to the divertor targets is modelled through the parallel losses). Finally, the \textit{wall shadow} starts at the end of the SOL and represents the part of the plasma in which the field lines are connected to solid surfaces other than the divertor targets. These could be the first wall or other limiter-like objects in the machine, which have the effect of greatly reducing the parallel connection length (and consequently increasing the parallel losses, see Eqs.\ref{8}-\ref{9} below). In the case of MAST this solid surface are the P5 coils, which are in vessel.    

In the presence of gradients of the thermodynamic fields, the edge region becomes unstable to interchange modes which generate turbulence. Advected by the turbulent fields, density and temperature fluctuations cross the LCFS, entering the SOL. Here, the interchange mechanism self-propels the blobs of plasma in the less dense and colder background \cite{Krashenninikov2001} until the excess of temperature and density is removed at the divertor plates or reaches the wall. 

It is appropriate, at this point, to note that our "turbulence engine" which creates the blobs is questionable. Indeed, electromagnetic drift wave turbulence, which is not captured in the ESEL model, is thought to be dominant in the edge of L-mode plasmas \cite{Scott1997,Scott2002}. On the other hand, the main subject of our study are the dynamics of the SOL, in which the interchange mechanism plays the most important role \cite{Ribeiro2005,Dudson2008}. It was observed in \cite{Dudson2008} that the filament motion in the SOL is not significantly affected by the "turbulence engine" used in the edge, thus supporting the philosophy of our approach. 

\subsection{Equations and dimensionless parameters}

The ESEL equations describe the evolution of the plasma density, $n$, electron temperature, $T$ and plasma vorticity, $\Omega$, which is the laplacian of the electrostatic potential, $\phi$:
\begin{eqnarray}
\label{1}
\frac{\partial n}{\partial t} +\frac{1}{B}[\phi,n] &=& nC( \phi) - C(nT_e) +D\nabla_\perp^2n -\Sigma_n n, \\
\label{2}
\frac{\partial T}{\partial t} +\frac{1}{B}[\phi,T] &=& \frac{2}{3}TC(\phi) - \frac{7}{3}TC(T)-\frac{2}{3}\frac{T^2}{n}C(n) +\chi\nabla_\perp^2T -\Sigma_T T, \\
\label{3}
\frac{\partial \Omega}{\partial t} +\frac{1}{B}[\phi,\Omega] &=& -C(nT_e)+\mu\nabla_\perp^2 \Omega -\Sigma_\Omega \Omega, \\
\label{4}
\Omega &=& \nabla_\perp^2 \phi.
\end{eqnarray}
To simplify the notation, we have that $B^{-1}[\phi,f]\equiv B^{-1}\textbf{b}\times\nabla\phi \cdot \nabla f$ (representing the advection of a generic field $f$ due to the $\textbf{E}\times\textbf{B}$ drift). $C(f) \equiv (\rho_s/R) \partial f/\partial y$ is the curvature operator, where $R$ is the major radius, $\rho_s$ is the ion Larmor radius calculated with the electron temperature and $y$ is the "poloidal" coordinate in the drift plane. The magnetic field, $B$ is assumed to vary as the inverse of the major radius, so that $B^{-1} \approx 1+ \epsilon +(\rho_s/R) x$, where $\epsilon$ is the inverse aspect ratio and $x$ is the "radial" coordinate (this expression does not contain an angular dependence since we are assuming proximity to the outer mid-plane). Note also that we have approximated the total magnetic field with the toroidal magnetic field, which could be sometimes questionable in spherical configurations since the poloidal magnetic field could be relevant at the edge. In the specific case that we discuss here, the measured $B_P$ at the outer midplane was $0.08$ T, which would increase the value we used by a mere 5\%. 

Equations \ref{1}-\ref{4} are Bohm normalized in order to highlight the relevant dimensionless parameters governing the problem. Besides the already introduced aspect ratio $\epsilon$ and interchange drive $\rho_s/R$, these include the perpendicular collisional dissipative coefficients (the particle diffusion, $D$, the heat conductivity, $\chi$, and the ion viscosity, $\mu$) and the inverse of the parallel loss times ($\Sigma_n$, $\Sigma_T$ and $\Sigma_\Omega$). The latter are used to represent the losses of particles, energy and momentum in the direction parallel to the confining magnetic field. Indeed, the model describes self-consistently only the perpendicular dynamics of the plasma fluctuations, i.e. their motion in the drift plane. 

The dimensionless parameters are calculated self-consistently from the plasma quantities measured in the experiment (e.g. edge density and temperature, safety factor, etc.). For $D$, $\chi$ and $\mu$ we use neoclassical collisional theory, while $\Sigma_n=\Sigma_\Omega$ are obtained assuming advective losses in the parallel direction and conductive losses are used to estimate $\Sigma_T$ \cite{Fundamenski2007,Militello2012}. In particular:
\begin{eqnarray}
\label{5}
D &=& (1+1.3q^2)(1+\theta)\frac{\rho_e^2\nu_{ei}}{\rho_s^2\Omega_i}\\
\label{6}
\chi &=& (1+1.6q^2)\left[4.66\frac{\rho_e^2\nu_{ee}}{\rho_s^2\Omega_i}+\Theta_{ie}2\frac{\rho_i^2\nu_{ii}}{\rho_s^2\Omega_i}\right]\\
\label{7}
\mu &=& (1+1.6q^2)\frac{3}{4}\frac{\rho_i^2\nu_{ii}}{\rho_s^2\Omega_i},
\end{eqnarray}
where $\rho_e$ and $\rho_i$ are the electron and ion Larmor radius, $\Omega_i$ is the ion gyration frequency, $\nu_{ss'}$ is the collision frequency between the species $s$ and the species $s'$ (with "i" for the ions and "e" for the electrons), $\theta=T_i/T_e$ and $\Theta_{ie} \equiv [1+(\nu_{e,\epsilon}^*/\nu^*_{e})^2]^{-1}$ with the equipartition collisionality $\nu_{e,\epsilon}^*\approx 63$ for a deuterium plasma and the collisionality $\nu_e^*\equiv L_\parallel/\lambda_e$ ($L_\parallel$ is the mid-plane to target connection length and $\lambda_e$ the electron collisional mean free path). Note that $q$ is the safety factor and the brackets that contain it represent the neoclassical correction to the dissipation coefficients. In addition:
\begin{eqnarray}
\label{8}
\Sigma_n &=& \Sigma_\Omega = \frac{M_\parallel \xi c_s}{l_\parallel\Omega_i} \\
\label{9}
\Sigma_T &=& \frac{2}{3}\frac{\chi_{\parallel,e}}{L_\parallel^2\Omega_i}, 
\end{eqnarray} 
where $M_\parallel$ is the Mach number of the parallel flows, $c_s \equiv \sqrt{T_e/m_i}$ where $m_i$ is the ion mass, $\xi\equiv \sqrt{Z+\theta}$ and $\chi_{\parallel,e}=3.2v_{te}^2/\nu_{ee}(1+4/\nu_e^*)^{-1}$ is the parallel electron heat conduction, which includes a flux limiter correction for low collisionalities ($v_{te}$ is the electron thermal velocity). In Eq.\ref{8}, the parameter $l_\parallel=L_\parallel/\alpha$ represents the length of the plasma filament in the parallel direction which, due to its ballooning nature, is a fraction of the total connection length ($\alpha$ is a constant, see below for further details). To be more precise, this model for the parallel losses is based on the realistic assumption that blobs emerge from the core at the outer midplane, as they are generated by the turbulence inside the separatrix. Such turbulence is ballooned as a result of the toroidal geometry. Therefore, as the filaments erupt in the SOL, they have a finite parallel extension (given by $l_\parallel$ and $L_\parallel$). Hence, in their initial evolution, the filaments are disconnected from the targets, which cannot influence their motion. 

It is worth noticing that since no neoclassical theory is rigorously derived in the absence of closed flux surfaces, Eqs.\ref{5}-\ref{7} must necessarily be considered approximate expressions. However, heuristic considerations on the physics of the Pfirsch-Schl\"uter flows, such as those contained in \cite{Fundamenski2007}, support their applicability even in the SOL. The interested reader can find a more complete and comprehensive discussion of the ESEL model and of the physics that it describes in \cite{Garcia2004,Fundamenski2007,Militello2012}.

The MAST discharge \#21712 was characterized by the edge quantities (measured at the LCFS) shown in Tab.\ref{tab1}. Using Eqs.\ref{5}-\ref{9} we obtain the dimensionless parameters needed for the ESEL simulation, see Tab.\ref{tab2} (for completeness, we give also the relevant electron collisionality, $\nu^*_e\approx 20$, which would correspond to the conduction limited regime). 

A few conclusive remarks on the model are appropriate here. First, the parallel connection length in a spherical tokamak like MAST cannot be calculated correctly by using the standard expression $L_\parallel \approx \pi q R$, which in our case would give $L_\parallel \approx 16.5$ m. The value $L_\parallel=10$m was obtained using field line tracing from the mid-plane to the target and is hence more realistic ($L_\parallel$ varies from 13m to 9m in the region between 6mm to 40mm from the LCFS, and $L_\parallel=10$m is an average over this region). 

Secondly, $\Sigma_n$ and $\Sigma_\Omega$ are estimated using $\alpha=2.5$ so that $l_\parallel \approx 4$m. This number was obtained from the observation that typical plasma filaments in MAST emerge from the outboard mid-plane and extend from upper to lower X-point \cite{Kirk2006,Kirk2007}. In other words, the parallel extension of the turbulent structures is longer than in conventional tokamaks, where a reasonable estimate would suggest $\alpha=6$. In addition, the value of the parallel loss terms changes radially in order to model the three different plasma regions. In the edge, $\Sigma_n=\Sigma_T=\Sigma_\Omega=0$ as no parallel loss is present in the confined region. The loss terms take their nominal values, i.e. those in Tab.\ref{tab2}, in the SOL, while in the wall shadow region this value is increased by a factor 20 in order to simulate the shorter connection length. 

Finally, the ion temperature at the separatrix was not available from experimental measurements in this specific discharge. However, in the proximity of the SOL this is typically a factor 2 higher than the electron temperature \cite{Elmore2012}, which is the value we chose.

\begin{table}
\caption{Edge quantities for MAST discharge \#21712 (measured at the LCFS except for $q_{95}$). Note that the edge magnetic field is obtained using: $B_{edge}=B_T(1+\epsilon)^{-1}$}
\label{tab1}
\begin{tabular}{|c|c|c|c|c|c|c|c|c|}
\hline  $n [10^{19}m^{-3}]$ & $T_{e} [eV]$ & $T_{i} [eV]$ &$B_{edge} [T]$ & $q_{95}$ & $M_\parallel$ & $L_\parallel [m]$ & $\epsilon$ & $R [m]$ \\ 
\hline  0.5 & 15 & 30 & 0.24 & 6.2 & 0.4 & 10 & 0.64 & 0.85 \\ 
\hline 
\end{tabular} \nonumber
\end{table}
\begin{table}
\caption{Dimensionless parameters for MAST discharge \#21712, corresponding to the dimensional values in Tab.\ref{tab1}.}
\label{tab2}
\begin{tabular}{|c|c|c|c|c|c|c|c|c|}
\hline  $D$  & $\chi$ & $\mu$ & $\Sigma_n=\Sigma_\Omega$ & $\Sigma_T$ & $\rho_s/R$  \\ 
\hline 0.0121  & 0.0440  & 0.1131  & $4.213 \cdot 10^{-4}$ & $1.864 \cdot 10^{-3}$ & $2.862 \cdot 10^{-3}$\\ 
\hline 
\end{tabular} \nonumber
\end{table}

\subsection{Numerical set up}

The code solves Eqs.\ref{1}-\ref{4} in a numerical box covering $150\rho_s$ in the "radial" direction and $75\rho_s$ in the "poloidal" direction. Since in our case $\rho_s=2.43$mm, this corresponds to simulating a region of the outer mid-plane of $36.45$cm times $18.23$cm. The edge, the SOL and the wall shadow regions each correspond to one third of the total box size in the radial direction (i.e. $50\rho_s\approx 12.15$cm each, which is the typical gap between LCFS and "wall" in MAST). The numerical grid uses $512 \times 256$ points so that $\Delta x=\Delta y=0.293 \rho_s$, thus assuring proper resolution for the turbulent structures. A shorter high resolution run ($1024\times 512$) was also performed to verify that grid size did not affect our results. In addition, the simulations are run well beyond the phase in which the turbulence reaches a statistical steady state. The end time is of the order of a few tens of $msec$ for MAST, which corresponds to several thousand turbulence correlation times. 

At the inner boundary $n$, $T$ are kept constant while $\Omega$ and $\phi$ are set to zero. At the outer boundary, we assume $\Omega=0$ and $\partial \phi/\partial x=v_y=0$ as well as zero temperature and density gradients, $\partial T/\partial x =\partial n/\partial x=0$. Finally, periodicity is imposed in the "poloidal" direction. 

\section{Comparison between experimental and numerical data} \label{SecIV}

In this Section we compare the experimental signal of the ion saturation current (analysed with the technique discussed in Sec.\ref{SecII}) with its counterpart obtained from the ESEL simulation, the parameters of which were described in Sec.\ref{SecIII}. 

\subsection{Average profiles}

As a first comparison, we examine the average profile of $I_{sat}\sim n\sqrt{T}$ in the SOL. In the following we represent this quantity with $<I_{sat}>$, where the angular brackets symbolize the average with respect to time and in the numerical case also with respect to the poloidal direction. $<I_{sat}>$ is therefore a function of the radial variable only and one single profile is representative of the whole ESEL simulation. For the experimental measurements, this function is calculated as the mean of the sub-signals at each radial position. 

\begin{figure}
\includegraphics[height=6.5cm]{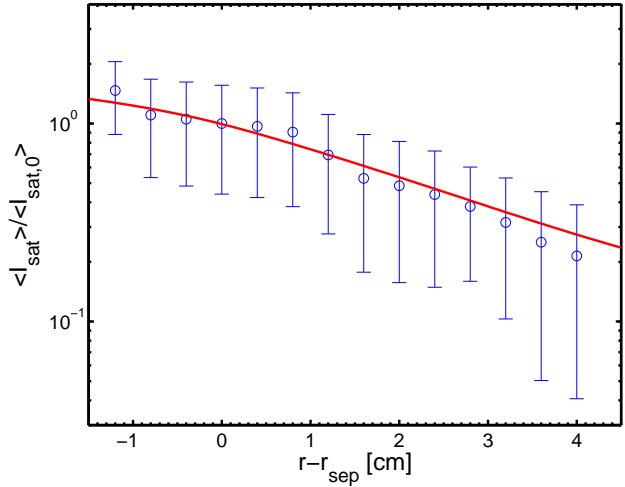} \caption{Ion saturation current normalized to its value at the LCFS for the experimental measurements (markers) and the ESEL simulation (solid line). The error bars represent one standard deviation of the experimental signal and display the level of the turbulent fluctuations.}
\label{fig1}
\end{figure}

Figure \ref{fig1} shows the average ion saturation current normalized with respect to its value at the LCFS for the experimental and the numerical case. The error bars associated with the experimental points represent one standard deviation of the signal and are used to display the level of the turbulent fluctuations (as such, they are not a measure of the intrinsic quality of the data). The agreement between the measurements and the numerical data is excellent, which suggests that the ESEL model is able to capture the key physical processes that determine the decay of the density and the temperature in the SOL. 

\begin{figure}
\includegraphics[height=6.5cm]{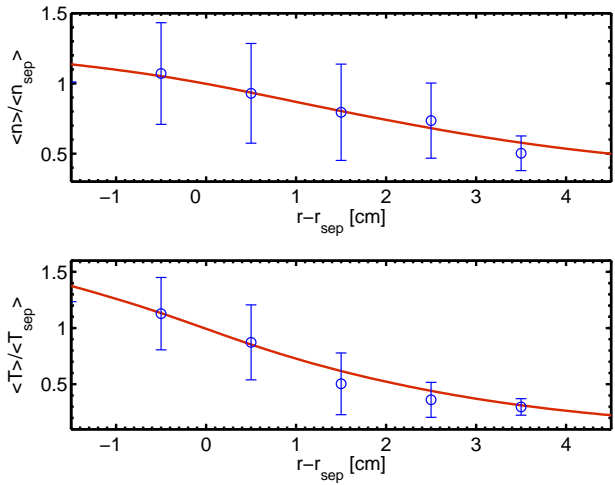} \caption{Plasma density (upper frame) and electron temperature (lower frame) normalized to their value at the LCFS for the Thomson scattering experimental measurements (markers) and the ESEL simulation (solid line). The error bars represent one standard deviation of the experimental signal.}
\label{fig1bis}
\end{figure}

In particular, the ESEL simulations allow us to estimate the density decay length, $\lambda_n =-<n>/(d<n>/dr)$, and temperature decay length, $\lambda_T = -<T>/(d<T>/dr)$, which for this particular discharge are around 6 cm and 3 cm respectively. This is confirmed by edge Thomson scattering measurements of average profiles of the density and electron temperature, which can be directly compared with the results of the simulations, see Fig.\ref{fig1bis}. Note that the Thomson scattering measurements have a coarser radial resolution, but also in this case the agreement with the simulations is good.  

\subsection{Statistics of the Fluctuations}

While the good agreement between experimental and numerical profiles is an indication of the appropriateness of the physical model used, an even more stringent test is the comparison of the statistics of the fluctuations of the signal. In this regard, several statistical quantities can be evaluated. 

\begin{figure}
\includegraphics[height=6.5cm]{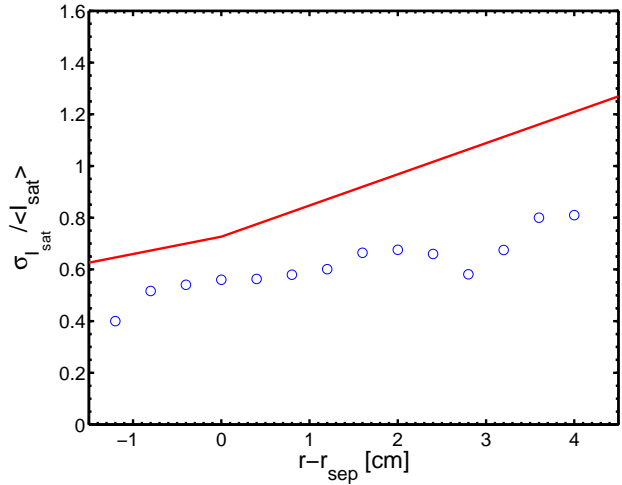} \caption{Relative amplitude of the fluctuations, $\sigma_I/<I_{sat}>$ for the experimental measurements (markers) and the ESEL simulation (solid line).}
\label{fig2}
\end{figure}

The most obvious is the amplitude of the fluctuations with respect to the mean. In our work, this quantity is calculated as the ratio between standard deviation of the signal, $\sigma_I$, and its mean value, $<I_{sat}>$ (in the sense described in the previous subsection).  The radial variation of the relative amplitude of the fluctuations is shown in Fig.\ref{fig2}.
In this case, the agreement is not as good as for the average profiles, with ESEL predicting larger fluctuations than those observed. On the other hand, the simulated values are not very far from the experimental measurements (20-30\% difference) and the code is able to capture the overall increasing trend as a function of radius. Note also that both the simulated and experimental results describe large fluctuations in the SOL, as commonly observed in most experimental machines. 

The skewness, $S$ and the kurtosis, $K$, are the third and the fourth standardized moment of the probability distribution function (PDF) of the signal and measure its asymmetry and peakedness. In the context of our problem, a positive skewness indicates a majority of bursts above the average, which are associated to hotter and denser plasma blobs. Conversely, a negative skewness is a sign of the predominance of plasma "holes", i.e. colder and less dense fluctuations in a hot and dense background. The kurtosis measures the relative importance of extreme events (i.e. events much larger or smaller than the average). Simply explained, a large kurtosis indicates that it is not uncommon in the SOL to encounter plasma filaments that are much hotter and denser than the background plasma. The so called flatness, $F\equiv K-3$, is a more convenient quantity than the kurtosis as $F=S=0$ for a Gaussian PDF.
\begin{figure}
\includegraphics[height=6.5cm]{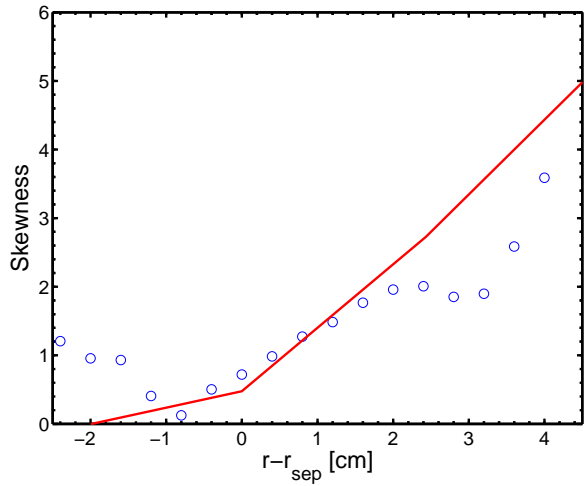} \caption{Skewness of the PDF of $I_{sat}$ for the experimental measurements (markers) and the ESEL simulation (solid line).}
\label{fig3}
\end{figure}
\begin{figure}
\includegraphics[height=6.5cm]{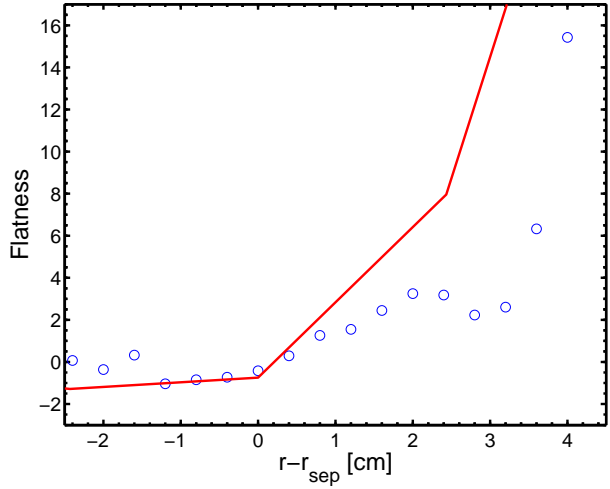} \caption{Flatness of the PDF of $I_{sat}$ for the experimental measurements (markers) and the ESEL simulation (solid line).}
\label{fig4}
\end{figure}

Figures \ref{fig3} and \ref{fig4} show the skewness and the flatness of the experimental and numerical signals. Good agreement is found in the region around the separatrix and in the near SOL. In the far SOL the ESEL fluctuations tend to be more skewed and with more extreme events than the experimental ones, although the overall increasing trend with major radius is correctly recovered.

\begin{figure}
\includegraphics[height=6.5cm]{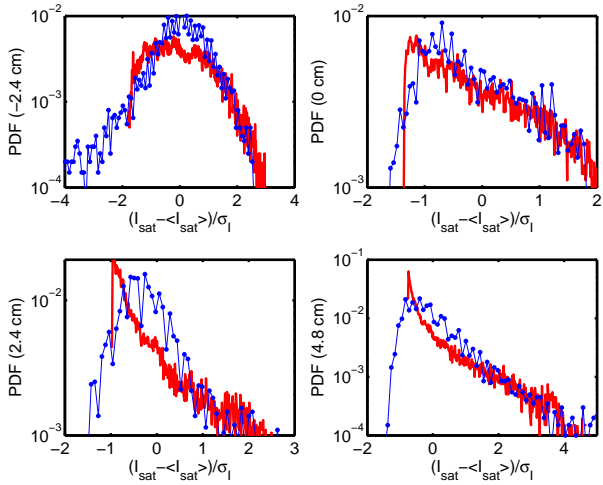} \caption{Probability Distribution Function of $I_{sat}$ calculated in four radial locations for the experimental measurements (thin line with markers) and the ESEL simulation (thick solid line).}
\label{fig5}
\end{figure}

Although very informative, the statistical moments provide only a limited picture of the fluctuation statistics and this can be sometimes misleading (as in the present case). For this reason, we turn now our attention to the full PDF of the fluctuations, calculated in four radial locations ($r-r_{sep}=-2.4,0,2.4,4.8$ cm). Figure \ref{fig5} shows a remarkable agreement between the measured and the simulated PDF. 
Note also the passage from quasi-Gaussian statistics inside the LCFS to a highly skewed and peaked PDF in the near and far SOL, an indication of the of the intermittent nature of the signal. It is interesting to note that the PDFs in the near and far SOL (at $r-r_{sep}=2.4$ cm and $r-r_{sep}=4.8$ cm) show less negative extreme events than their experimental counterpart. In other words, our model properly captures the statistics of the blobs, while it underestimates the magnitude of the plasma "holes". This is affecting the calculation of the statistical moments and explains the discrepancy of $S$ and $F$ in Figs.\ref{fig3}-\ref{fig4}. Note that the PDF at $r-r_{sep}=-2.4$ cm is shown only to demonstrate that our interchange edge turbulence has a Gaussian statistics. However, we expect drift waves to dominate the actual turbulence, so that the good agreement is likely accidental. In view of all these results, we can argue that the ESEL model provides a good description of the fluctuation statistics as well as of the average profiles.   

\subsection{Shape and dynamics of the filaments}    

We complete our comparison by analysing some morphological properties of the bursty events in the $I_{sat}$ signal which are generated as a plasma filament passes by the position of the probe. To determine the typical time evolution of the bursts we calculate conditionally averaged temporal wave forms with a trigger condition based on the maximum amplitude of the fluctuation: $(I_{sat}-<I_{sat}>)_{max}/\sigma_I>2.5$. At $r-r_{sep}=2.4$ cm, this criterion selects 52 bursts in the numerical signal and 71 in the experimental signal, while at $r-r_{sep}=4.8$ cm the bursts are 71 and 42 respectively.  
\begin{figure}
\includegraphics[height=6.5cm]{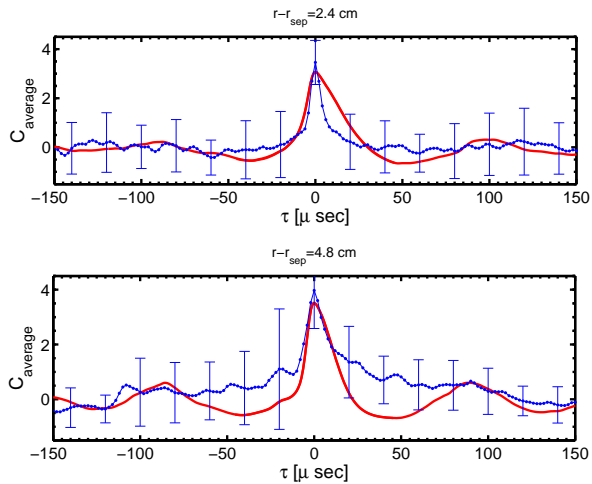} \caption{Conditionally averaged wave form of $I_{sat}$ at two radial locations for the experimental signal (thin line with markers) and the ESEL simulation (thick solid line). In the figure, $C_{average}\equiv (I-<I>)/\sigma_I$ for $(I-<I>)_{max}/\sigma_I>2.5$. The error bars represent one standard deviation in the experimental data.}
\label{fig6}
\end{figure}

The conditionally averaged wave forms are shown in Fig.\ref{fig6} and suggest that the typical blob structure in MAST is quite different from other tokamaks. In particular, no clear steep front and trailing wake are present in our experimental measurements, although these are commonly observed in other machines \cite{Sanchez2000,Boedo2001,Antar2003,Garcia2006}. These results are consistent with previous analysis of MAST fluctuations \cite{Antar2003}. It therefore appears that the front of the MAST filaments is more spread out and does not show a sharp transition. This could be caused by the fact that in the SOL of MAST the collisional diffusion in the perpendicular direction is particularly large and it acts to smooths out the blob (see e.g. \cite{Militello2012} for a comparison with TCV). Indeed, the perpendicular particle diffusion scales like $\rho_e^2\nu_{ei}$ (see Eq.\ref{5}) and in MAST $\rho_e$ is much larger than in conventional tokamaks due to the much smaller confining magnetic field (for similar edge temperature). The ESEL simulations capture this peculiar behaviour and agree reasonably well in both the shape and maximum amplitude of the conditionally averaged wave form.

\begin{figure}
\includegraphics[height=6.5cm]{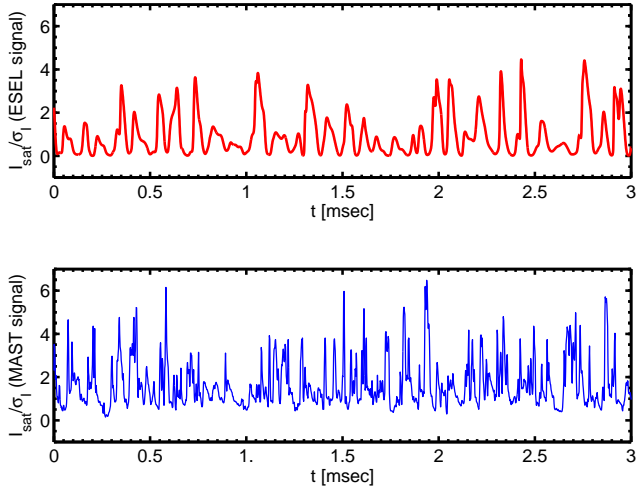} \caption{Comparison between experimental (lower frame) and simulated (upper frame) time signals of the normalized ion saturation current, $I_{sat}/\sigma_I$ evaluated at $r-r_{sep}=2.4$cm.}
\label{fig7}
\end{figure}

On a more global scale, the comparison between the time traces of the experimental and simulated $I_{sat}/\sigma_I$ signals is shown in Figs.\ref{fig7} for $r-r_{sep}=2.4$ cm.
Although the amplitude, the duration and the interval between burst events is comparable in the two cases, the ESEL signal is smoother and fails to capture the small scale fluctuations (compare also the bottom left frame in Fig.\ref{fig5}). This could be due to by the absence of kinetic effects or drift instabilities in the fluid simulation, which could contribute to fragment the filament. Other causes of this discrepancy could be the presence of fast transitory events in the parallel directions or simply noise in the experimental signal. A similar lack of small scale fluctuations was previously found in BOUT simulations of MAST filaments \cite{Dudson2007}. In order to make sure that these short scale fluctuations did not affect the conclusions of our analysis, we filtered out all the frequencies above 20kHz from the experimental signal and we compared the new PDFs with those shown in Fig.\ref{fig5}. We found only minor variations, in particular slightly lower PDFs for values of $(I_{sat}-<I_{sat}>)/\sigma_I$ close to zero, which confirmed the excellent agreement with the numerical data (they even improved it in the case $r-r_{sep}=2.4$ cm).   

\begin{figure}
\includegraphics[height=6.5cm]{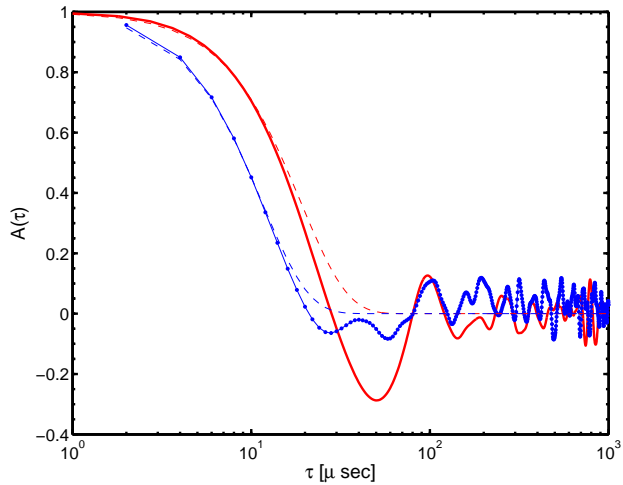} \caption{Comparison between experimental (thin line with markers) and simulated (thick solid line) autocorrelation functions of the normalized ion saturation current, evaluated at $r-r_{sep}=2.4$cm. The dashed lines represent the exponential fit obtained using Eq.\ref{10}.}
\label{fig8}
\end{figure}

A more quantitative way to characterise the temporal scales present in the $I_{sat}$ signal is to use the autocorrelation function of the fluctuations, $A(\tau)\equiv E[\widetilde{I}_{sat}(t)\widetilde{I}_{sat}(t+\tau)]/\sigma_I^2$, where $\widetilde{I}_{sat}\equiv I_{sat}-<I_{sat}>$ and "$E$" is the expected value operator. Figure \ref{fig8} shows $A(\tau)$ for the radial position $r-r_{sep}=2.4$ cm.
For both the experimental and the numerical data, we fitted the autocorrelation function with the expression:
\begin{equation}
\label{10}
A_{fit}(\tau)=\exp[-(\tau/\tau_c)^{\beta-1}],
\end{equation}
where the autocorrelation time, $\tau_c$, and the so called cascade index, $\beta$, are fitting parameters \cite{Graves2005}. In this context, the autocorrelation time represents the characteristic length of the burst events, which is related to typical time scale of the plasma filaments. The cascade index is related to the ideal scale-free power spectrum, $P(f)\sim f^{-\beta}$, where $f$ is the frequency. Note, however, that our analysis is not able to provide details of the cascade mechanism or even to prove its existence, as this would require more sophisticated statistical tools \cite{Manz2008}. In this context, therefore, $\beta$ should be seen as a parameter describing the shape of the autocorrelation function.

The best fit of the experimental data is obtained with $\tau_c=11.5$ $\mu$sec and $\beta=2.66$, while the best fit of the ESEL results gives $\tau_c=19$ $\mu$sec and $\beta=2.66$ (see dashed lines in Fig.\ref{fig8}). The same analysis repeated at different radial position led to similar discrepancies in $\tau_c$. It is also interesting to notice that the first positive maximum of the two autocorrelation functions, roughly representing the interval between two successive bursts, occurs at the same time, $\tau \approx 100$ $\mu$sec. This is further confirmed by counting the number of events above one standard deviation per unit time, which are around 9 per msec for the numerical signal and 14 per msec for the experimental one (the average is performed on longer time traces than those shown in Fig.\ref{fig7}). In addition, we calculated the power spectra for both signals at this radial position and found that they peak at a value close to 10 kHz (not shown), which indicates a dominant time scale of 100 $\mu$sec. 

To conclude this Section, we discuss the 2D structure of the plasma filaments in the drift-plane observed in the ESEL simulation. Unfortunately, in this case no direct comparison with the reciprocating probe data is possible. On the other hand, MAST is equipped with fast CCD cameras which allow a systematic characterisation of the $D_\alpha$ emission coming from the filaments. This diagnostic provided an estimate of their perpendicular width in L-mode discharges, which is around 5-10 cm \cite{Dudson2008}.    
\begin{figure}
\includegraphics[height=6.5cm]{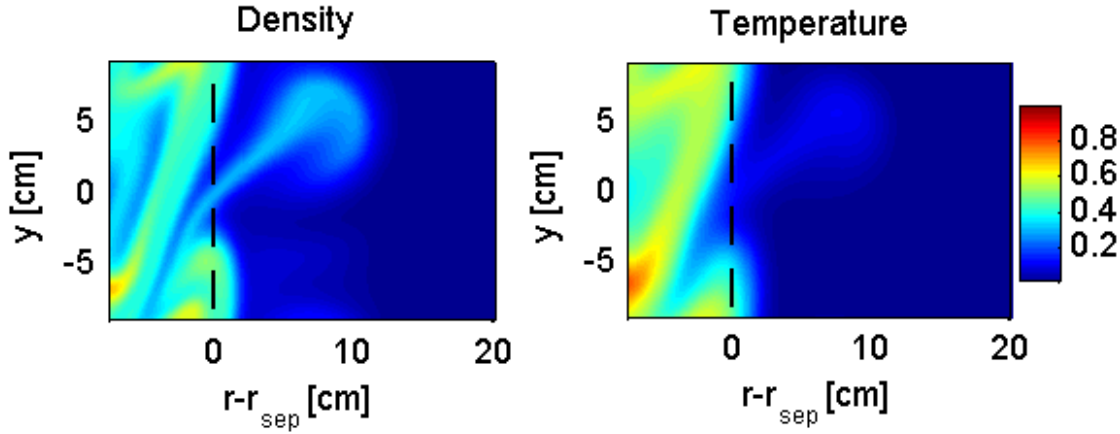} \caption{Density (left frame) and electron temperature (right frame) in the drift-plane. The figures capture the eruption of a plasma filament.}
\label{fig9}
\end{figure}
Figure \ref{fig9} displays typical features of the simulated plasma filaments, which erupt continuously when the turbulence is in a statistical steady state. We first note that the blob has a width that is largely compatible with the experimental observations. In addition, its boundaries are blurred and the mushroom structures observed in numerical simulations of other machines \cite{Garcia2006,Fundamenski2007} are only vaguely recognisable. Together with the smooth conditionally averaged wave form, this is another sign of the peculiar structure of the filaments in MAST. 

\section{Effect of the temperature fluctuations on the measurement of the potential} \label{SecV}

In the previous Section, we successfully validated our numerical model against the data of the reciprocating probe. The next step is to use the detailed physical insight that the ESEL simulations can offer in order to better interpret the experimental results. We describe here an application of the numerical modelling, which is used to assess the quality of the plasma electric potential measurements in MAST. 

As mentioned in Section \ref{SecII}, the Langmuir probes installed on the reciprocating arm measure the ion saturation current as well as the floating electrostatic potential, $\phi_{fl}$. The latter depends on the actual electrostatic plasma potential, $\phi$, through the following expression \cite{StangbyBOOK}:
\begin{equation}
\label{11}
\phi_{fl}=\phi+\frac{1}{2}T_e\log\left[2\pi\kappa^2\left(1+\frac{T_i}{T_e}\right)\frac{m_e}{m_i}\right],
\end{equation}
where $\kappa$ is the ratio between the ion and electron probe collecting areas, $m_e$ and $m_i$ are the electron and ion mass. It is a standard approximation to take $\kappa\approx 1$ and $T_i/T_e \approx 1$ so that for a deuterium plasma Eq.\ref{11} reduces to: 
\begin{equation}
\label{12}
\phi_{fl}\approx\phi-2.83T_e.
\end{equation}    

Despite the simplicity of this expression, an accurate calculation of the plasma potential is often elusive since electron temperature measurement in the SOL are particularly difficult. This is due to the fact that this quantity is often obtained through the $I-V$ characteristic of the probe, which requires a sweep of the probe potential. When available, such sweeps are done at frequencies that are much slower than that of the turbulence, so that the short time scale variation of the temperature is lost. For this reason, it is common practise to identify the fluctuations of the plasma potential with those of the floating potential, $\widetilde{\phi}_{fl}\approx \widetilde{\phi}$. With sweeping probes, Eq.\ref{12} can be used, although with unreliable statistics for $T_e$. Crucially, the plasma potential is instrumental in the calculation of the $\textbf{E}\times\textbf{B}$ velocity and ultimately in the evaluation of the cross-field particle and energy fluxes in the SOL. Consequently, an imprecise estimate of $\phi$ cascades down to these important indirect measurement. An alternative but less common strategy is to use advanced probes, such as ball-pen probes which can provide a fast and localized measurement of the temperature. The experimental observations based on this approach seem to suggest that the temperature fluctuations cannot be neglected in Eq.\ref{12} (see e.g. \cite{Horacek2010}, where some of the results are interpreted with ESEL).

While experimental measurements of $T_e$ are not straightforward, the electron temperature is an output of our numerical simulations. It is therefore easy to assess the level of error introduced by neglecting the temperature fluctuations in Eq.\ref{12}. 
\begin{figure}
\includegraphics[height=6.5cm]{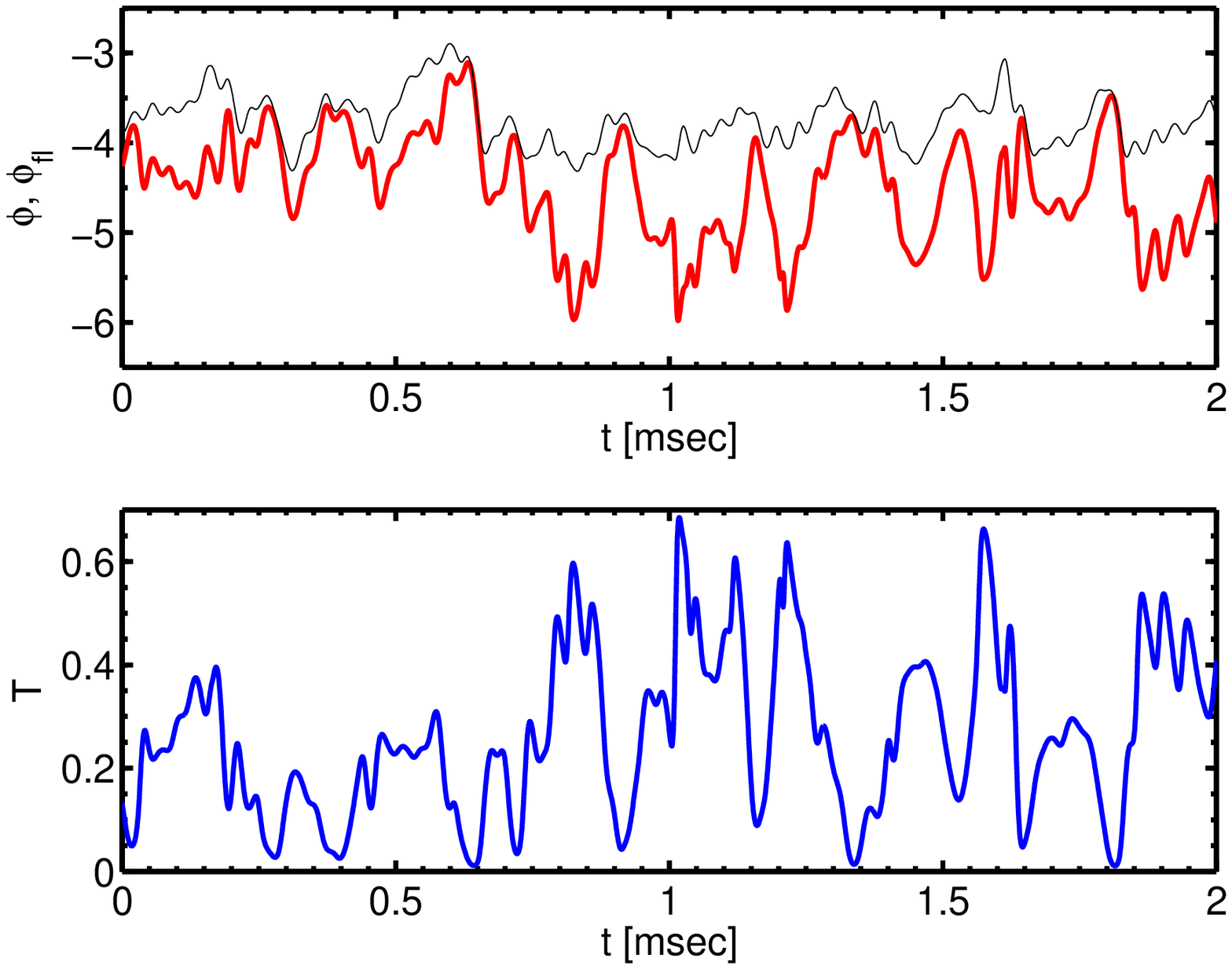} \caption{Upper frame: comparison between floating (thick line) and plasma (thin line) potential at the last closed flux surface. Lower frame: temperature fluctuations associated with the signal. Potentials and temperature are normalized as in Eqs.\ref{1}-\ref{4}. } 
\label{fig10}
\end{figure}
\begin{figure}
\includegraphics[height=6.5cm]{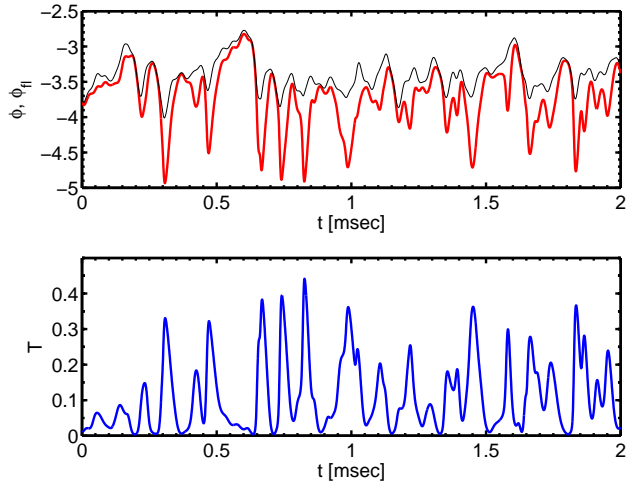} \caption{Upper frame: comparison between floating (thick line) and plasma (thin line) potential at $r-r_{sep}=2.4$ cm. Lower frame: temperature fluctuations associated with the signal. Potentials and temperature are normalized as in Eqs.\ref{1}-\ref{4}.} 
\label{fig11}
\end{figure}
Figure \ref{fig10} shows the numerical time traces of the floating and plasma potentials (upper frame) and the electron temperature fluctuations (lower frame) recorded at $y=0$ and $r-r_{sep}=0$ (see Fig.\ref{fig9} for a reference). Figure \ref{fig11} shows the same single point measurements for $y=0$ and $r-r_{sep}=2.4$ cm. While $\phi$ and $T$ are outputs of the code, $\phi_{fl}$ is calculated using Eq.\ref{12}. Both figures show time traces 2 msec long, extracted from signals lasting approximatively 20 msec. 

The floating and plasma potential differ at the LCFS as well as in the SOL, more markedly in the first case. Note that a more realistic choice of $T_i/T_e\approx 2$ would lead to a similar disagreement since the coefficient multiplying the temperature fluctuations in Eq.\ref{12} would pass from $2.83$ to $2.63$. From a quantitative point of view, it is useful to introduce the correlation coefficient, $\mathcal{R}_{\phi\phi_{fl}}\equiv E[(\phi_{fl}-\overline{\phi}_{fl})(\phi-\overline{\phi})]/(\sigma_{\phi_{fl}}\sigma_{\phi})$, which determines the linear dependence between the two curves as well as if one signal is representative of the other. Here, the overline symbols and the $\sigma$s represent the means and the standard deviations of the signals. The correlation coefficient can take values between $0$ and $1$, corresponding to no correlation or perfect correlation respectively. At the LCFS $\mathcal{R}_{\phi\phi_{fl}}=0.72$ while at $r-r_{sep}=2.4$ cm we have $\mathcal{R}_{\phi\phi_{fl}}=0.83$ (these numbers are calculated using the full 20 msec long signals).   

As mentioned above, the main reason for measuring the electrostatic potential is to calculate the velocity in the SOL. In particular, with the normalization used, we have that the dimensionless radial $\textbf{E}\times\textbf{B}$ velocity is simply $V_r=\partial \phi/\partial y$. When this quantity is experimentally evaluated there are at least two sources of errors. As discussed already, the first is related to the fact that the probes measure the floating instead of the plasma potential. The second is the finite separation between the Langmuir probes, which "transforms" the derivative of the potential into a finite difference. This implies that the measured velocity is not a true local quantity and is inaccurate if the potential field varies on scales smaller than the separation of the probes. The Gundestrup probe which collected the data we analysed in the previous three Sections evaluates the $\textbf{E}\times\textbf{B}$ velocity using two pins separated by 2.39 cm (see \cite{Higgins2012} for a schematic of the probe).   

We now compare the actual plasma velocity, which is calculated numerically in our simulations, with two synthetic measurements designed to evaluate the errors discussed above. In order to study the effect of the finite probe separation in MAST, we calculate $V_{r,\Delta y}\equiv \Delta \phi/\Delta y$. Here $\Delta \phi$ is the difference between the plasma potential measured by synthetic probes at $y=1.195$ cm and $y=-1.195$ cm so that their poloidal separation is $\Delta y = 2.39$ cm (this is repeated at different radial positions). In addition, we also evaluate $V_{r,fl}\equiv \Delta \phi_{fl}/\Delta y$, which combines the error due to the finite separation with the fact that they are calculating the floating instead of the plasma potential.  
\begin{figure}
\includegraphics[height=6.5cm]{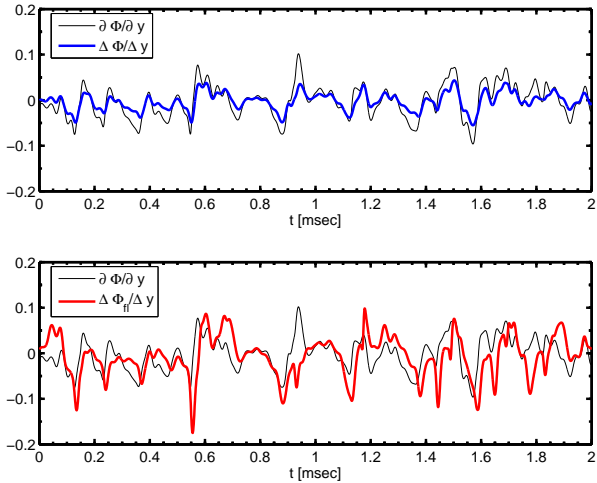} \caption{Upper frame: comparison between the actual radial velocity, $V_r=\partial \phi/\partial y$ (thin line) and the radial velocity calculated with the plasma potential using two probes, $V_{r,\Delta\phi}=\Delta \phi/\Delta y$ (thick line). Lower frame: comparison between the actual radial velocity, $V_r=\partial \phi/\partial y$ (thin line) and the radial velocity calculated with the floating potential using two probes, $V_{r,fl}=\Delta \phi_{fl}/\Delta y$ (thick line). All the signals are calculated at the LCFS at the poloidal position $y=0$.} 
\label{fig12}
\end{figure}
The curves produced by this procedure are shown in Fig.\ref{fig12}. The upper frame, which compares $V_r$ (thin line) with $V_{r,\Delta y}$ (thick line), suggests that the probe separation induces an error, although a limited one, as the correlation coefficient is $\mathcal{R}_{V_r V_{r,\Delta y}} = 0.91$. It is interesting to note that the synthetic measurement captures the pattern of the actual velocity but it fails to reproduce the amplitude of the maxima and the minima. Conversely, the $V_{r,fl}$ signal poorly represents the plasma velocity as their correlation coefficient is $\mathcal{R}_{V_rV_{r,fl}}=0.35$. Moving radially outward the correlation between the $\textbf{E}\times\textbf{B}$ velocity and $V_{r,\Delta y}$ remains around 0.9, while the agreement with $V_{r,fl}$ improves (but it remains poor) as a consequence of the fact that the temperature fluctuations become smaller and smaller. In particular, $\mathcal{R}_{V_rV_{r,fl}}=0.44$ at $r-r_{sep}=2.4$ cm and $\mathcal{R}_{V_rV_{r,fl}}=0.47$ at $r-r_{sep}=4.8$ cm. 
\begin{figure}
\includegraphics[height=6.5cm]{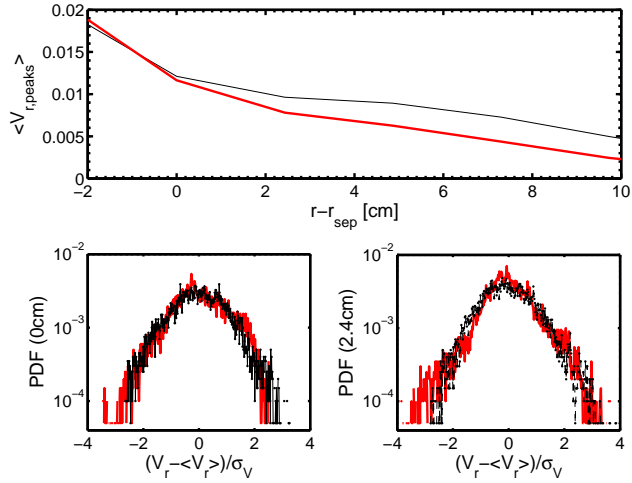} \caption{Upper frame: comparison between the time average profile of the peak radial velocity for the real signal (thin line) and the one calculated with the synthetic probe (thick line). Lower frames: comparison between the PDFs of the fluctuations of the radial velocity calculated from the real signal (thin line with markers) and the synthetic one (thick line). The PDFs evaluated at the LCFS and at $r-r_{sep}=2.4$ cm are evaluated in the left and right frame, respectively.} 
\label{fig13}
\end{figure}

Often the experimental measurements focus on the peak radial velocities, which are associated with the filaments. We identify them as the maxima of the signals in the lower frame of Fig.\ref{fig12}. Interestingly, the average value of these velocity peaks calculated with the $V_{r,fl}$ signal is not far from that obtained with $V_r$, although it always underestimates it in the SOL (see the upper frame of Fig.\ref{fig13}). In addition, also the statistics of the fluctuations is similar, as the lower frames in Fig.\ref{fig13} show. This suggests that the probe measurements might be used to provide useful information on the actual plasma velocity, despite the bad correlation with the actual signal. Whether this is true also in different experimental conditions is left for future work.

\begin{figure}
\includegraphics[height=6.5cm]{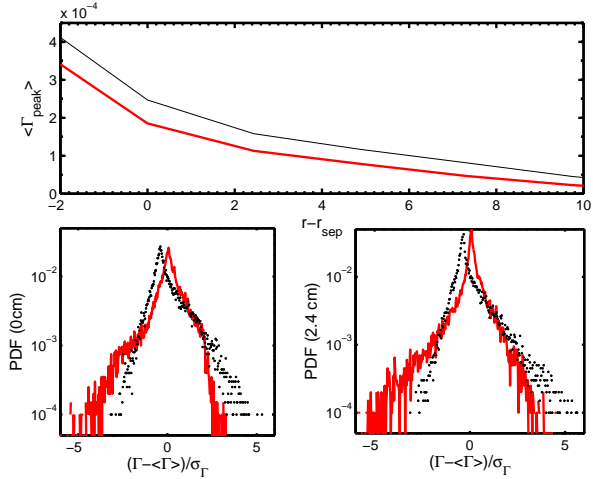} \caption{Same as Fig.\ref{fig13} for the radial particle flux and its fluctuations.} 
\label{fig14}
\end{figure}
We conclude by briefly discussing the reliability of the radial particle flux resulting from the synthetic probe. In order to calculate this quantity, we multiply the actual or the synthetic velocities by the plasma density (we therefore do not discuss how the temperature fluctuations affect the measurement of $n$ from $I_{sat}$). Also in this case, we consistently find that the synthetic measurement underestimates the peak particle flux, as seen in the upper frame of Fig.\ref{fig14}. On the other hand, the discrepancy is of the order of 25\% of the actual value, so that measured peak particle flux is the right order of magnitude. The lower frames of Fig.\ref{fig14} show the comparison between the PDFs of $nV_r$ and $nV_{r,fl}$ at two radial positions in the SOL. In both cases the agreement is poor, which implies that probe signals cannot provide reliable estimates of the statistics of the radial particle flux.  

\section{Summary and Conclusions}\label{SecVI}

Numerical simulations of the MAST Scrape-Off Layer, performed with the ESEL code, were compared with reciprocating probe measurement obtained during the reference L-mode discharge \#21712. This is the first time that a thorough validation of the code and of the model is attempted for a spherical tokamak. The dimensionless parameters employed in the simulations were calculated self-consistently using neoclassical transport theory (for $D$, $\chi$ and $\mu$) or reasonable assumptions for the parallel losses (advective losses for $\Sigma_n$ and $\Sigma_\mu$ and conductive losses for $\Sigma_T$). It is important to remark that in this process no free parameter was introduced and tuned to match the experimental data. 

The results thus obtained suggest that, despite their simplicity, Eqs.\ref{1}-\ref{4} are able to capture several features of the dynamics of the SOL. In particular, the time averaged and statistical properties of the ion saturation current, as well as the perpendicular size of the filaments, are quantitatively reproduced in the simulations. A remarkable agreement is found in the averaged profile of $T_e$, $n$ and $I_{sat}$ and in the PDFs of the fluctuations of the latter at different radial positions. The shape and length of the bursts as well as the interval between them are qualitatively captured, although the matching between numerical and experimental data is not as impressive. This confirms the success of the interchange paradigm for the evolution of the filaments in the SOL, although it suggests that the modelling requires refinements. In particular, while the simulations can reproduce the characteristic smooth shape of the bursts in MAST, they are not completely successful in replicating the small time scale fluctuations. Kinetic and electromagnetic effects \cite{Militello2011}, interaction with drift waves and a better description of the parallel dynamics \cite{Havlickova2011a, Havlickova2011b} might improve the agreement. The inclusion of these mechanisms in the model and the study of their effect is left for future work. 

In addition, Eqs.\ref{1}-\ref{4} are derived under several restrictive assumptions, such as a simplified magnetic geometry, an elementary description of the parallel losses and the total absence of neutral particles. The overall agreement with the experimental measurements (obtained also in TCV \cite{Garcia2006} and JET \cite{Fundamenski2007}) seems to indicate that in L-mode these effects only determine the details of the perpendicular transport.  

As an application, we used our simulations to determine the effect of the temperature fluctuations on Langmuir probe measurements of the plasma potential and radial velocity, a problem that has recently attracted the attention of the community \cite{Gennrich2012,Nold2012}. In addition, we also studied how the finite poloidal separation of the probes affects the calculation of $V_r$, a topic discussed also in \cite{Kirk2011}. Our results show that these combined effects (but mostly the first) contribute to decorrelate the measured and the actual signal. Although the situation improves moving outwards in the SOL, the correlation factor of the velocity remains well below 0.5. In other words, an accurate estimate of $\phi$ and $V_r$ require precise measurement of the floating potential and \textit{also} of the electron temperature. In particular, conditional averages triggered by the measured velocity signal (or applied to it) might lead to significantly wrong conclusions. On the other hand, in the case treated, both the average peak velocity and particle flux are qualitatively captured by the synthetic probe but consistently underestimated.     

\section{Acknowledgements}

F.M. gratefully acknowledges helpful discussions with Dr. S. Allan, Dr. G. Fishpool, Dr. J. Harrison and Dr. W. Morris. This work was funded by the RCUK Energy Programme under grant EP/I501045 and the European Communities under the contract of Association between EURATOM and CCFE. The views and opinions expressed herein do not necessarily reflect those of the European Commission.

\end{document}